# Programmable radio-frequency calculations in electromagnetic-wave domain

Shao Nan Chen[1,†], Zhan Ye Chen[1,2,†], Si Ran Wang[3†], Bi Rui[1], Jin Feng Kang[1], Zheng Xing Wang[3], Zhen Jie Qi[1], Lijie Wu[1], Hui Dong Li[1], Jun Yan Dai[1,*], Qiang Cheng[1,*] and Tie Jun Cui[1,2,*]

[1] State Key Laboratory of Millimeter Waves, Southeast University, Nanjing 210096, China

[2] Institute of Electromagnetic Space, Southeast University, Nanjing 210096, China

[3] State Key Laboratory of Terahertz and Millimeter Waves, City University of Hong Kong, Hong Kong, 999077, China

[†]Equally contributed to this work

[*]E-mail: junyand@seu.edu.cn, qiangcheng@seu.edu.cn, and tjcui@seu.edu.cn

**Abstract**

Information metasurfaces have emerged as pivotal components in next-generation electronic systems, with significant progress in their applications to communication, radar, and sensing. However, the current researches are mainly focused on their physical structures and system functions, while radio-frequency (RF) signal processing and calculation remain constrained to digital-domain operations. This reliance on digital conversion inherently increases hardware complexity and power consumption. To address this challenge, we propose a programmable RF calculation system based on a space-time-coding metasurface (STCM), which can control the wave-matter interactions through space-time-coding (STC) strategies and achieve direct RF calculations in the electromagnetic (EM) space in a reprogrammable way. Particularly, the fundamental signal operations - Fourier transform and convolution - are implemented in the EM-wave domain successfully. We validate the RF calculation capabilities in radar scenarios, facilitating the accurate detection of target velocity and range. Theoretical analysis, numerical simulations, and experimental results collectively demonstrate that the STCM-based RF calculation system exhibits superior precision, enhanced operational efficiency, and notable cost-effectiveness, highlighting its significant potentials for the next-generation electronic system deployments.

# Introduction

Over the past decade, metasurfaces have driven significant advances in electromagnetic (EM) physics, providing breakthrough technologies to manipulate EM fields and waves. These planar structures are made up of subwavelength artificial units that can precisely control the properties of the incident wave, such as the amplitude, phase, polarization, and propagation direction, in order to meet a wide range of application requirements[1–6]. In 2014, Cui *et al.* proposed the concept of digital coding and programmable metasurfaces[7], which unlocked new potentials by adopting a digital information perspective. This approach involves establishing mathematical mappings between the EM properties and spacial coding sequences of digital states of the metasurface elements. Information metasurfaces were further presented by integrating the EM wavefront control with the information processing through the use of space-time-coding (STC) strategies[8–13]. This advancement has enabled the development of novel electronic and information systems, including dynamic beam scanning[7,11,14], wireless communications[8,15–19], radar waveform syntheses[20,21], and intelligent sensing platforms[10,22,23].

Despite the breakthrough progress of information metasurfaces in designing new systems, most studies have primarily been focused on innovations in their physical structures and function designs, with limited progress in direct radio-frequency (RF) signal processing and calculations. The current RF signal processing still relies on digital signal processors (DSPs) in backend baseband modules. The analog-to-digital/digital-to-analog conversion (ADC/DAC) modules in the signal chain inevitably lead to the issues such as hardware redundancy, energy consumption, and transmission latency[24–27]. To address these limitations, the concept of RF calculation has emerged and garnered extensive attention[28–30]. As an emerging calculating paradigm, the core of RF calculation lies in using RF signals simultaneously as information carriers and calculating entities[31–33]. By performing the signal processing directly at physical layer, it significantly reduces the need for cross-domain conversion. Thus the RF calculation is regarded as a key development direction for the future backend information processing, holding important theoretical significance and engineering value to improve the energy efficiency and real-time performance of electronic information systems.

While the RF calculation offers significant theoretical promise and engineering value, the existing implementations are limited in the circuit domain and lack efficient architectures that are cost-effective, high-precision, and multifunctional simultaneously[31]. For instance, the conventional mixers perform spectral translation via dot-product operations between the local oscillator and input signals — a physical realization of RF domain multiplication. However, the mixers rely on complex circuit implementations, and their functional scope is constrained by single frequency-shifting operations, making them inadequate for sophisticated calculation tasks. More importantly, the traditional architecture is difficult to be reprogrammable. By contrast, the RF calculation devices based on information metasurfaces are expected to demonstrate greater research and application potentials. The time-domain coding strategy introduces new degrees of freedom to metasurface functional design[34], enabling information metasurfaces to be further refined into space-time-coding metasurface (STCM)[36–40]. STCMs integrate multidimensional EM wave manipulations with direct signal processing on the metasurface, thereby eliminating the need for complex RF backend circuits and demonstrating strong potentials for complex calculations. Moreover, the programmable nature of STCM makes it possible to realize programmable RF calculations. As the application scope has been expanded from the traditional EM regulation to signal processing, STCM is therefore highly compatible with the information processing requirements of the RF calculation, making it a suitable implementation approach for the RF calculation.

Here, we propose a programmable RF calculation system based on STCM, and as an example, physically implement two fundamental signal processing algorithms in EM-wave domain: the Fourier transform and convolution operation. The Fourier transform enables the frequency-domain signal analysis[41,42], while the convolution operation characterizes system responses to input signal[43,44], both constituting the essential elements for advanced signal processing. Our RF calculation architecture is based on rigorous theoretical derivation, numerical simulation, and experimental validation. The measured results agree well with the theoretical predictions, thereby confirming the advancement and feasibility of the proposed architecture. As an application, the system is successfully used in radar scenarios for range and velocity detection. The expanded information-processing capabilities and highlighted engineering utility make the STCM system a novel enabling tool for the programmable RF calculations in the wave domain.

## Theoretical framework of RF calculations by STCM

The implementation process of the STCM-based programmable RF calculation system is illustrated in Fig. 1. When the RF signal to be calculated is incident on STCM, the RF calculations will be directly performed on STCM by dynamically regulating its reflection characteristics in a programmable way. Without losing the generality, when a signal with carrier frequency $f_c$ and baseband signal $I(t) = |A(t)| e^{j\varphi_i(t)}$ impinges on the STCM, the reflected signal

$E_r(t)$ can be modeled as:

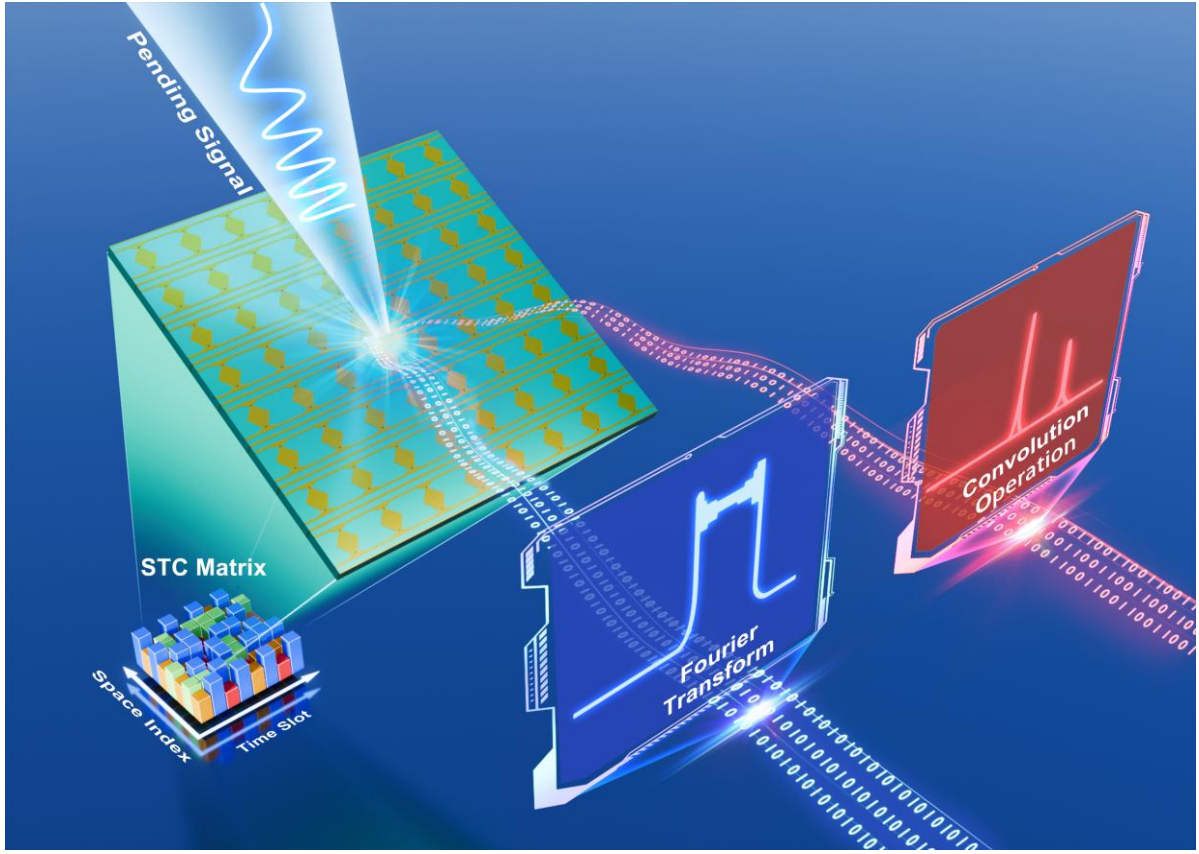

**Fig. 1**. The STCM-enabled programmable RF calculation system. When a signal impinges upon STCM, STCM will reflect the outcomes subjected to diverse calculation processes in a programmable way according to the coding strategies corresponding to distinct STC matrices.

$$E_r(t) = |A(t)\Gamma(t)| e^{j\{2\pi f_c t + \varphi_i(t) + \varphi_r(t)\}} \tag{1}$$

where $|A(t)|$ and $|\Gamma(t)|$ represent the amplitude of the incident signal and metasurface reflection coefficient, respectively, and $\varphi_i(t)$ and $\varphi_r(t)$ are the corresponding phase terms. Then the baseband component of the reflected signal is extracted as:

$$R(t) = |A(t)\Gamma(t)| e^{j[\varphi_i(t)+\varphi_r(t)]} = |A(t)| e^{j\varphi_i(t)} \cdot |\Gamma(t)| e^{j\varphi_r(t)} = I(t) \cdot \Gamma(t) \tag{2}$$

This equation provides a strict multiplicative relation among the reflected signal, the incident signal, and the reflection coefficient of the metasurface. Since the core idea of information metasurface is to manipulate EM waves by designing digital coding sequences on metasurface atoms, the continuous signal in **Eq. (2)** can be discretized into vector form, thus constructing a discretized time modulation model for the RF signal:

$$\begin{cases} \mathbf{I} = [I[0], I[1], \cdots, I[N-1]]^T \in \mathbb{C}^{N \times 1} \\ \mathbf{\Gamma} = [\Gamma[0], \Gamma[1], \cdots, \Gamma[N-1]] \in \mathbb{C}^{1 \times N} \end{cases} \tag{3}$$

where *N* is the number of uniform time-domain samples per signal period. By superimposing and integrating the signal energy at discrete time points, the total system response is expressed as:

$$Z = \Sigma \mathbf{R} = \mathbf{\Gamma} \cdot \mathbf{I} \tag{4}$$

where $\Sigma$ represents element-wise vector summation. This model reveals the mechanism of the STCM's RF-information modulation and provides a mathematical foundation for the RF calculations in the EM-wave domain, instead of circuit domain.

The programmable RF calculation system achieves signal processing functions through dynamic manipulations of EM parameters. Its mathematical essence can be reduced to solve a linear operation of the following form:

$$y[k] = \sum_{n=0}^{N-1} x[n] \cdot h[k,n], \ k = 0, \cdots, K-1 \tag{5}$$

where *n* and *k* represent the time indices of the system's input and output vectors, while *N* and *K* denote the lengths of the corresponding vectors. The system's output vector $\mathbf{Y} \in \mathbb{C}^{K \times 1}$ can be regarded as the interaction between the system input $\mathbf{X} \in \mathbb{C}^{N \times 1}$ and the system response $\mathbf{H} \in \mathbb{C}^{K \times N}$. In the programmable RF calculation system shown in Fig. 1, the input signal corresponds to the discrete time-domain sampling sequence $\mathbf{I} \in \mathbb{C}^{N \times 1}$ of the incident wave, while the system response is characterized by the reflection coefficient matrix $\mathbf{\Gamma} \in \mathbb{C}^{K \times N}$. To establish a mathematical model equivalent to Eq. (5), the reflection coefficient matrix $\mathbf{\Gamma}$ is decomposed into a combination of row vectors:

$$\mathbf{\Gamma} = [\mathbf{\Gamma}_0, \mathbf{\Gamma}_1, \cdots, \mathbf{\Gamma}_k, \cdots, \mathbf{\Gamma}_{K-1}]^T \tag{6}$$

$$\mathbf{\Gamma}_k = [\Gamma_k[0], \Gamma_k[1], \cdots, \Gamma_k[N-1]] \tag{7}$$

where $\mathbf{I}$ is a periodic function with the period of *N* and repetition count of *K*, while $\mathbf{\Gamma}_k$ represents the reflection coefficient encoding vector for the *k*-th cycle. The system output at cycle *k* can be expressed as:

$$Z[k] = \Sigma \mathbf{R}_k = \sum_{n=0}^{N-1} \mathbf{I}[n] \cdot \mathbf{\Gamma}_k[n] \tag{8}$$

The design of the STCM coding matrix $\mathbf{\Gamma}$ directly determines the response of the calculation system. To intuitively establish the mapping relationship between the system response and the coding units, Eq. (5) is expanded into a matrix form:

$$Z_{K\times 1} = \begin{bmatrix} \Sigma\mathbf{R}_0 \\ \Sigma\mathbf{R}_1 \\ \vdots \\ \Sigma\mathbf{R}_{K-1} \end{bmatrix} = \begin{bmatrix} \Gamma_0[0] & \Gamma_0[1] & \cdots & \Gamma_0[N-1] \\ \Gamma_1[0] & \Gamma_1[1] & \cdots & \Gamma_1[N-1] \\ \vdots & \vdots & \ddots & \vdots \\ \Gamma_{K-1}[0] & \Gamma_{K-1}[1] & \cdots & \Gamma_{K-1}[N-1] \end{bmatrix} \times \begin{bmatrix} I[0] \\ I[1] \\ \vdots \\ I[N-1] \end{bmatrix} \quad (9)$$

Substituting into **Eq.(5)**, the output response $Z \in \mathbb{C}^{K\times 1}$ of the system is expressed as:

$$\mathbf{Z} = [\Gamma_0 \cdot \mathbf{I}, \ \Gamma_1 \cdot \mathbf{I}, \cdots, \ \Gamma_{K-1} \cdot \mathbf{I}]^T \quad (10)$$

This model demonstrates that STCM can achieve the linear operation in the EM-wave domain when it satisfies specific conditions. In the following discussions, we will conduct theoretical derivations and numerical simulations by constructing the coding matrix $\Gamma$ for the Fourier transform and convolution operation, combined with the discrete time-domain signal model **I**.

## Fourier transform

The Fourier transform serves as a fundamental mathematical tool in modern electronic systems. By converting signals between time and frequency domains, they provide critical theoretical support for system design and optimization[41]. In the STCM-based programmable RF calculation system, its digital coding characteristics require the use of the discrete Fourier transform (DFT) to realize the physical layer operation in the EM-wave domain. For a discrete signal with $N$ sampling points, Eq. (5) can be modified to the following solution paradigm[45,46]:

$$y[k] = \sum_{n=0}^{N-1} x[n] \cdot e^{-j\frac{2\pi}{N}kn}, \quad k = 0, 1, \cdots K-1 \quad (11)$$

where $y$ and $x$ represent the outputand input signals, respectively, with $K$ and $N$ denoting the lengths of the output and input signal vectors. To preserve the orthogonal completeness of the Fourier transform, the length of the output sequence must satisfy $K = N$. The computational mechanism of the STCM lies in mapping the mathematical operation of the Fourier transform onto a physical modulation process. By dynamically modulating the reflection phase, the STCM synthesizes the complete set of basis functions $e^{-j\frac{2\pi}{N}kn}$ of the Fourier transform. When the incident signal interacts with the STCM, its time-varying reflection coefficient directly apply the corresponding frequency weighting associated with each basis function to the signal. Based on this design principle, the reflection coefficient matrix is constructed as:

$$\Gamma_{K\times N} = \begin{bmatrix} 1 & 1 & \cdots & 1 \\ 1 & \omega^{-1\cdot 1} & \cdots & \omega^{-1\cdot(N-1)} \\ \vdots & \vdots & \ddots & \vdots \\ 1 & \omega^{-(K-1)\cdot 1} & \cdots & \omega^{-(K-1)(N-1)} \end{bmatrix}, \omega = e^{j\frac{2\pi}{N}} \quad (12)$$

where the $k$-th row of $\Gamma_{K\times N}$ represents the reflection coefficient of the STCM at the $k$-th time slot. During this slot, the interaction between the incident wave and STCM directly produces the Fourier coefficient of the *k*-th frequency component in the EM-wave domain. By performing element-wise multiplication between the incident signal and the time-varying reflection coefficient across all slots, followed by accumulation, the frequency spectrum of the incident signal is obtained.

For example, we apply the Fourier transform to a 5kHz monochromatic sinusoidal signal using STCM. The sampling frequency and duration are set to 20 kHz and 1.6 ms, respectively, resulting in $N = 32$ sampling points, as shown in Fig. 2(a). When the signal illuminates STCM, its reflection coefficient is modulated to compute the Fourier coefficients over the 32 time slots. Fig. 2(b) shows the reflection phases of the STCM for the first four slots, and the resulting spectrum in Fig. 2(c) closely matches the theoretical result.

As an RF calculation system, STCM provides fully programmable control over its system parameters. We further configure the STCM to double the effective sampling duration with the fixed sampling rate, thereby yielding 64 sampling points. As illustrated in Fig. 2(d), this adjustment extends the sampling window for the same 5kHz input signal, enabling STCM to compute a larger set of Fourier coefficients. The corresponding reflection phase profiles for the first four slots are shown in Fig. 2(e). Compared with Fig. 2(b), the phase waveforms remain unchanged, but their modulation periods are doubled, leading to a halved modulation frequency of the underlying basis functions and an enhanced frequency resolution. The resulting spectrum in Fig. 2(f) shows excellent agreement with the theory. The transition from Figs. 2(a-c) to 2(d-f) demonstrates that STCM can reconfigure the parameters of the Fourier transform operations in real time, highlighting its intrinsic programmable capability.

## Convolution operations

Here we discuss the implementation of circular convolution in the EM-wave domain using STCM. Circular convolution is a convolution operation[47] defined on finite-length or periodic sequences, which has been widely applied in OFDM communication systems, radar signal processing, and image filtering[43,48]. Specifically, the circular convolution can be expressed as:

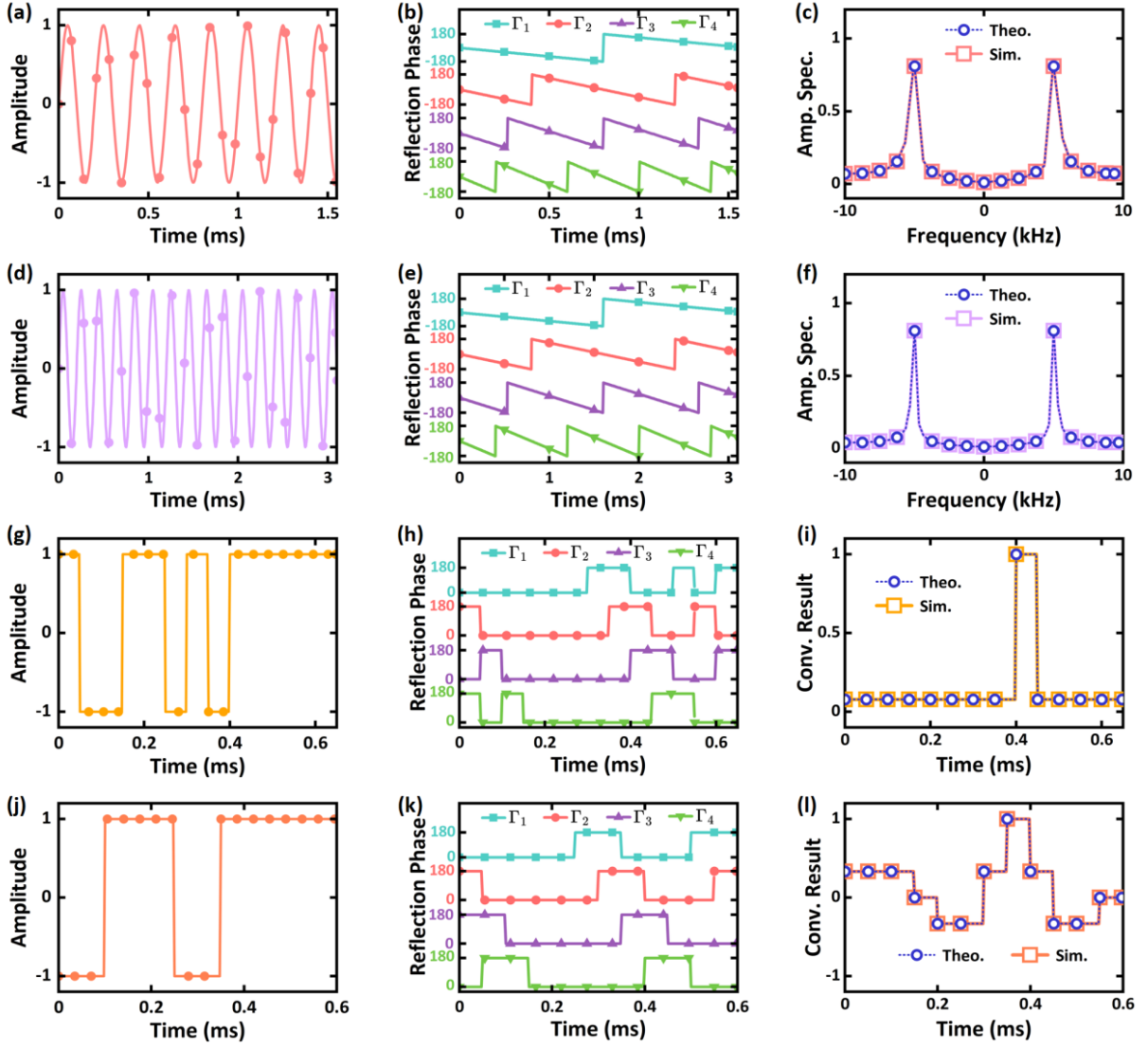

**Fig. 2. Simulation results of STCM-based RF calculations. (a-f) Fourier transform operations:** (a, d) Input signal amplitudes for the 5 kHz monochromatic sinusoidal signal under 32-point and 64-point (doubled window) sampling, respectively. (b, e) Corresponding partial STCM reflection phase sequences (first four time slots) for each case. (c, f) Fourier transform results, comparing theoretical values (blue circles) with simulations (red and purple squares). **(g-l) Convolution operations for matched filtering:** (g, j) Input signal amplitudes of the 13-bit Barker code and the 12-bit Hadamard sequence, respectively (20 kHz sampling, 50 μs symbol duration). (h, k) STCM phase sequences of the corresponding matched-filter kernels (first four time slots). (i, l) Convolution outputs, comparing theoretical values (blue circles) with simulations (yellow and orange squares).

$$y[k] = \sum_{n=0}^{N-1} x[n] \cdot h\big[(k-n) \bmod N\big], k = 0, 1, \cdots K-1 \quad (13)$$

in which $y$, $x$, and $h$ respectively denote the output signal, input signal, and the convolution kernel, and $N$ and $K$ represent the lengths of the input and output signal vectors. In the circular convolution, both input and output vectors have the same length, i.e., $K = N$, because the convolution is performed under periodic boundary conditions. With time-varying reflection coefficient, STCM can construct the convolution kernel in the time dimension. Specifically, the reflection coefficient of the STCM can be given by:

$$\mathbf{\Gamma}_{K\times N} = \begin{bmatrix} h[0 \bmod N] & h[(-1) \bmod N] & \cdots & h[(1-N) \bmod N] \\ h[1 \bmod N] & h[0 \bmod N] & \cdots & h[(2-N) \bmod N] \\ \vdots & \vdots & \ddots & \vdots \\ h[(K-1) \bmod N] & h[(K-2) \bmod N] & \cdots & h[(K-N) \bmod N] \end{bmatrix} \quad (14)$$

where each row of the matrix represents the STCM reflection coefficient modulated at a specific time slot. The reflection coefficient sequence for the $k$-th time slot corresponds to the convolution kernel $h[n]$ after time reversal and a circular shift by $k$ samples. During this slot, the interaction between the incident wave and the STCM executes the $k$-th step of the sliding-window operation of discrete convolution. Consequently, by performing the inner product between the input signal and the shifted kernel directly in the EM-wave domain, the $k$-th output sample $y[k]$ is obtained.

The programmability of STCM enables the on-demand implementation of application-specific convolution kernels, such as those used in matched filtering. In the matched-filter operation[49], the required convolution kernel depends on the input waveform. As illustrated in Figs. 2(g) and 2(j), the incident signals are a 13-bit Barker code and a 12-bit Hadamard sequence, respectively. The time duration of each symbol in the transmitted sequence is set to $T_c = 50 \ \mu s$, and the sampling frequency of the simulated signal is configured as $f_s = 20 \ \text{kHz}$. The STCM is programmed to implement two distinct convolution kernels, configured respectively as the time-reversed 13-bit Barker code and the 12-bit Hadamard sequence, as shown in Figs. 2(h) and 2(k). As a result, the output signals are shown in Figs. 2(i) and 2(l), where the theoretical values are denoted by blue dashed lines for comparison. The simulated convolution results exhibit excellent agreement with these theoretical predictions, verifying the effectiveness and accuracy of the STCM-enabled RF matched filtering operation. The matched-filter output of the 13-bit Barker code exhibits a dominant main lobe with extremely low sidelobes, reflecting its strong sidelobe suppression. In contrast, the 12-bit Hadamard sequence produces higher sidelobes, consistent with its orthogonality-oriented design. These results agree well with the theory, validating the STCM-enabled convolution operation, and further demonstrate that the programmable STCM allows real-time reconfiguration of the convolution kernel to accommodate different incident signals.

## Experimental validations

Based on the theoretical derivations and numerical simulations, a series of experiments were conducted to validate the practical performance of the STCM-based programmable RF calculation system. Firstly, an 8×12 2-bit reflective STCM sample was fabricated, in which each column shares the same control signal. As shown in Fig. 3(a), the metasurface adopts a three-layer structure: the top layer is a patch layer composed of a hexagonal patch and DC feed lines, the middle layer is a F4B dielectric substrate ($\varepsilon_r = 2.65, \tan\delta = 0.001$), and the bottom layer is a metal ground plane. The specific geometry parameters are listed as follows: $P = 20$ mm, $H = 5$ mm, $L_1 = 6.62$ mm, $L_2 = 6.25$ mm, $L_3 = 3.8$ mm, $L_4 = 8$ mm, and $L_5 = 12.82$ mm. The hexagonal patch is electrically interconnected with the feed lines via PIN diodes mounted on both sides, enabling the generation of four phase states (00/01/10/11) through controlled on/off combinations of two diodes. The measured reflection amplitude and phase responses are depicted in Figs. 3(b) and (c), respectively. Experimental results show that, at 5 GHz, the metasurface exhibits discrete phase responses of 0°, 90°, 180°, and 270° while maintaining a reflection amplitude above -3 dB. To enhance the performance under oblique incidence, the dielectric substrate is integrated with two columns of metallized vias aligned along the orthogonal polarization direction. Measured data confirms that, over an incidence angle range of ±40°, the metasurface maintains stable reflection characteristics without significant performance degradation. Obviously, merely regulating the four reflection states of STCM is insufficient to meet the requirements of the programmable RF calculation system for the reflection coefficient in the preceding theoretical analysis. In Supplementary Information 1, we propose a method to construct the equivalent reflection coefficient $\Gamma_{equ}$ by means of spatial encoding. This modulation strategy enables conventional metasurfaces to meet the modulation requirements of RF calculations.

For the designed metasurface sample, we plot the In-phase & Quadrature (IQ) distribution of $\Gamma_{equ}$ in Fig. 3(d), in which the colored scatter points within the red circle ($|\Gamma_{equ}| \leq A_{ref}$) correspond to arbitrary amplitude and phase that can effectively meet the requirements of the RF calculations for the reflection coefficient. The detailed modulation protocol and implementation results are presented in Supplementary Information 2.

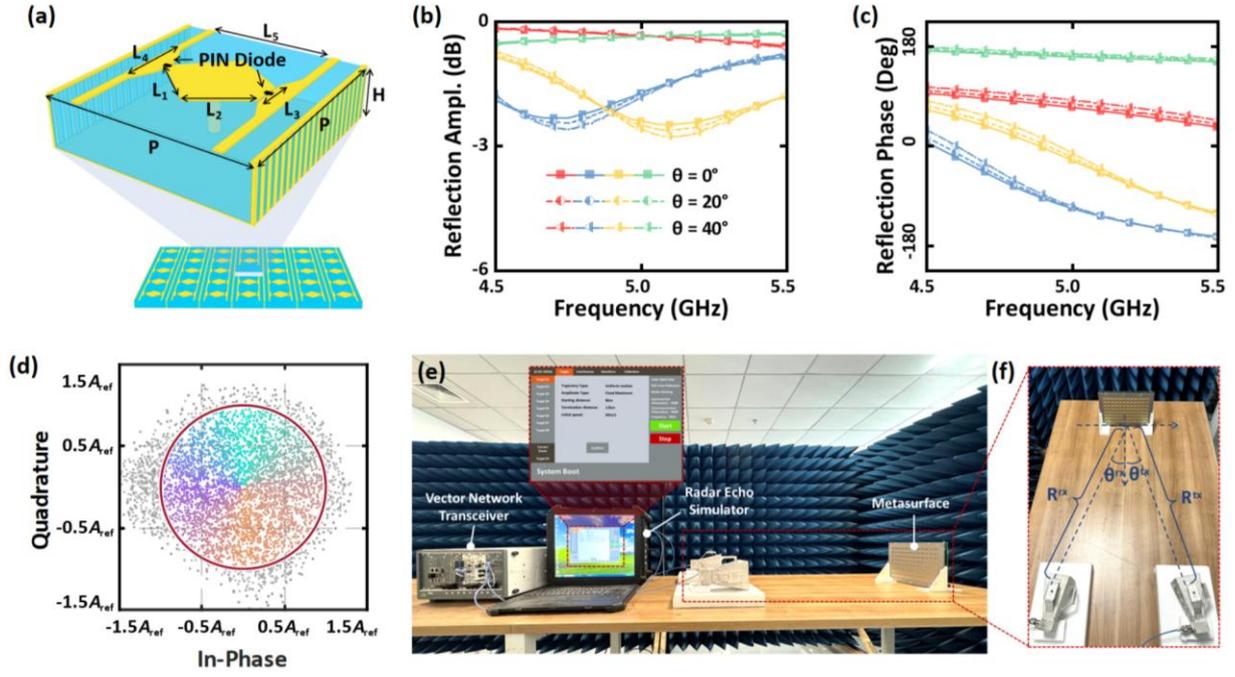

**Fig. 3. Hardware implementation and characterization of the programmable STCM-based RF calculation system.** (a) Schematic of the tunable meta-atom enabling 2-bit phase control. (b, c) Measured reflection (b) amplitude and (c) phase responses of the metasurface under incidence angles up to 40° at 5 GHz. (d) (d) Experimentally modulated equivalent reflection coefficient $\Gamma_{equ}$, demonstrating extended phase-amplitude control via spatial coding. (e) Architecture of the programmable RF calculation system. (f) Spatial layout of the measurement setup: transmitting and receiving horn antennas positioned at spherical coordinates (1 m, 20°, 0°) and (1 m, –20°, 0°), respectively, relative to the STCM center.

Subsequently, we constructed a programmable RF calculation system with the STCM, as depicted in Figs. 3(e) and 3(f). The system was composed of two modules: the data transceiver module and the spatiotemporal modulation module. The data transceiver module was a vector signal transceiver (PXIe-5841) connected to two pyramidal horn antennas, enabling signal transmission to and reception from the RF calculation system. The spatiotemporal modulation module was composed of STCM and an FPGA controller, which was used to generate the reflection coefficient matrix $\Gamma$. In the experimental setup, the vector signal transceiver generated a 5-GHz RF signal for calculation, which was subsequently radiated onto the STCM. Concurrently, the FPGA generated a control voltage waveform and loaded it onto the STCM, whereby the reflection coefficient of the metasurface was adjusted in accordance with the predefined coding strategies. The vector signal transceiver then captured the STCM's reflected echoes and retrieved the RF calculation results. Here, we construct two experiments: STCM-based Fourier transform and convolution operation.

For the Fourier transform demonstration, we selected three representative signal types, as summarized in Table 1: linear frequency modulation (LFM), single-tone, and multi-tone composite. The number of points for the Fourier transform was set to $N = 512$, which defined the number of samples in the transform and directly determined the frequency resolution of the output. In the experiment, the signal sampling rate was set to $f_s = 100$ kHz. To verify whether the STCM can accurately modulate the target reflection coefficient required for performing the Fourier transform operation, we first tested the reflection coefficient coding sequences used for RF calculation. According to Eq. (12), the modulated reflection coefficient should maintain constant amplitude while their phase varies over time at a specific frequency. The test results for four time slots ($\Gamma_1 \sim \Gamma_4$) are shown in Figs. 4(a)-(d), where the scatter points represent the measured reflection coefficient codes and the solid red curves denote the theoretical values. The *x–y* plane of the coordinate system corresponds to the IQ distribution of the reflection coefficient, and the *z*-axis represents time. Although minor deviations are observed in some scatter points, the evolution trend of the vast majority of the data agrees well with the theoretical prediction, thereby confirming the effectiveness of implementing the Fourier transform operation based on the STCM. These occasional deviations may originate from limited control signal precision and environmental noise interference, whose impact on the overall computation results is negligible. Supplementary Information 3 further provides some additional test data of the STCM reflection coefficient waveforms for other time slots, offering further support for the above conclusion.

For the LFM signal input, the phase and amplitude spectra calculated by the STCM are shown in Figs. 4(e) and 4(f). A comparison between the measured data (solid curves) and theoretical baselines (dashed curves) shows that the RF calculation results agree well in the main energy-concentrated regions. Some deviations are confined to low-energy zones, primarily due to spatial clutter interference. However, the spectral energy in these regions has decayed to the background noise floor, so they do not affect the core feature extraction of the Fourier transform. Furthermore, the calculated frequency-domain result is transformed back to the time domain, as illustrated in Fig. 4(g). The reconstructed time-domain signals closely match the theoretical waveforms of the input signals, demonstrating that the RF calculation faithfully preserves the signal's spectral characteristics.

For the single-tone input, the time-domain waveform is displayed in Fig. 4(h). Fig. 4(i) presents the theoretical spectrum and the spectrum produced by STCM. Despite minor noise, the measured spectrum agrees well with the theory, demonstrating that STCM can directly realize the Fourier transform in the EM-wave domain. Consistent with the theoretical results in Figs. 2(g)-(l), reducing the number of transform points will decrease the effective sampling window and degrade spectral resolution. Benefiting from the system's programmability, the Fourier transform length can be reconfigured without modifying the hardware. As illustrated in Fig. 4(j), a 64-point transform is obtained in the EM-wave domain by adjusting the sampling duration. Although the spectral resolution is reduced, the resulting spectrum remains in good agreement with the digital reference.

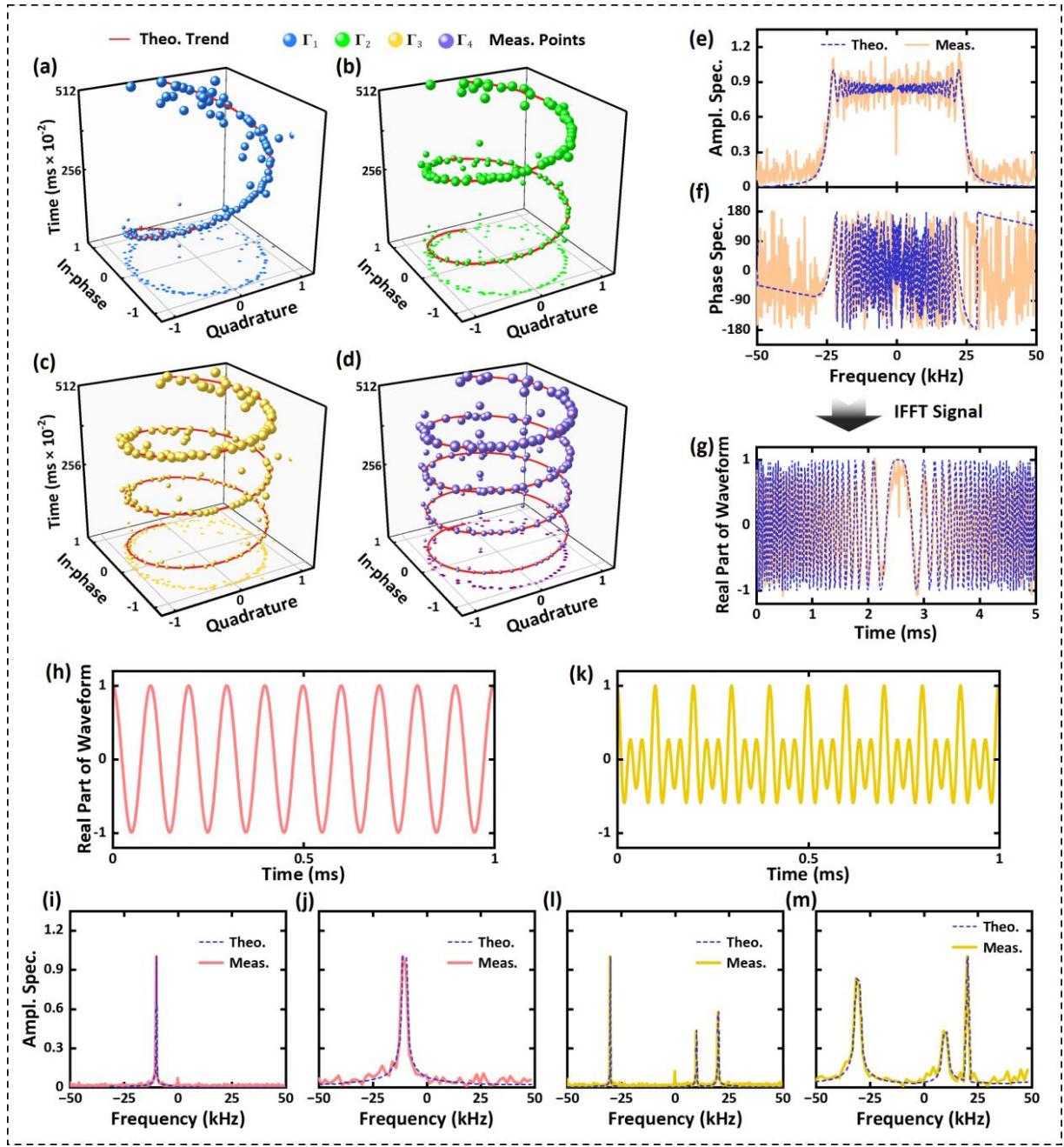

**Fig. 4. Fourier transform outcomes of the programmable STCM-based RF calculation.** (a-d) Partial reflection coefficient coding sequences ($\Gamma_1 \sim \Gamma_4$) used for wavefront modulation. (e–g) RF calculation results for LFM signal (512 points): (e) amplitude spectrum, (f) phase spectrum, (g) signal reconstructed via IFFT. (h–m) Results for single-tone and multi-tone composite signals: (h) real part of incident single-tone waveform; (i) corresponding RF-calculation result (512 points); (j) corresponding RF-calculation result (64 points); (k) real part of incident

multi-tone composite waveform; (l) corresponding RF-calculation result (512 points); (m) corresponding RF-calculation result (64 points).

Furthermore, because the Fourier transform does not require switching among multiple system responses, a unified set of reflection coefficient codes is sufficient to implement the transform. This property allows the same STCM coding sequence to handle arbitrary input signals. To verify this, we introduced a multi-tone composite waveform (Fig. 4(k)) while maintaining the same coding sequence for both the 512-point and 64-point configurations. The resulting spectra (Figs. 4(l) and 4(m)) match the theoretical predictions, further confirming the correctness of the proposed EM-wave Fourier transform framework.

Table 1. Mathematical models and parameter specifications of four testing waveforms for the Fourier transform.

| Case | Signal type | Mathematical model | Parameters |
| --- | --- | --- | --- |
| [1] | LFM signal | $x_1[n] = \exp\left\{j\pi \cdot \left[2nf_1/f_s + K_r\left(n/f_s\right)^2\right]\right\}$ | $f_1 = -25$ kHz<br>$K_r = 9.77$ MHz/s |
| [2] | Single-tone signal | $x_2[n] = \exp\{j \cdot 2\pi nf_2/f_s\}$ | $f_2 = -10$ kHz |
| [3] | Multi-tone composite signal | $x_3[n] = \sum_{i=3}^{5} A_i \exp\{j \cdot 2\pi nf_i/f_s\}$ | $[f_i] = [-30, 10, 20]$ kHz<br>$[A_i] = [4, 2, 3], i = 3, 4, 5$ |

For the STCM-based convolution demonstration, we chose two representative cases: low-pass filtering and matched filtering. For the low-pass filtering case, the STCM was modulated to implement a Hann-windowed sinc kernel with a prescribed cut-off frequency (Case 1 in Table 2). A close agreement can be observed between the theoretical kernel (Fig. 5(a)) and its measured STCM-modulated counterpart (Fig. 5(b)), with only minor deviations arising from control-signal fluctuations and ambient noise, confirming accurate kernel synthesis in the EM-wave domain. Then, we constructed an input signal $x_4[n]$ consisting of two frequency components ($f_6 = 10$ kHz and $f_7 = 35$ kHz (Fig. 5(c)). As shown in Fig. 5(d), the reflected waveform produced by the STCM agrees well with the numerically calculated convolution result. In the frequency domain, as illustrated in Figs. 5(e)-(f), the high-frequency component at $f_7$ is strongly attenuated, whereas the low-frequency component at $f_6$ is retained, therefore achieving the expected low-pass response.

For the matched filtering case, STCM was modulated to construct the time-reversed complex conjugate of the target signal (Case 2 in Table 2) as the convolution kernel for maximizing the output signal-to-noise ratio (SNR) at the frequency of interest. To verify the matched-filter kernels generated by the STCM, we compared the measured STCM reflection coefficient waveform with the theoretical waveform, as shown in Figs. 5(g) and 5(h), with high consistency observed in their real parts. Building on this, an input signal (Fig. 5(i)) consisting of a tone at 10 kHz embedded in 0 dB-SNR pseudo-random noise was generated and used to illuminate the STCM. The real part of the reflected waveform closely matches the theoretical value (dashed line), as shown in Fig. 5(j). In the frequency domain (Figs. 5(k)-(l)), the component at 10 kHz is significantly amplified, while the in-band noise is effectively suppressed, demonstrating the intended matched-filtering behavior.

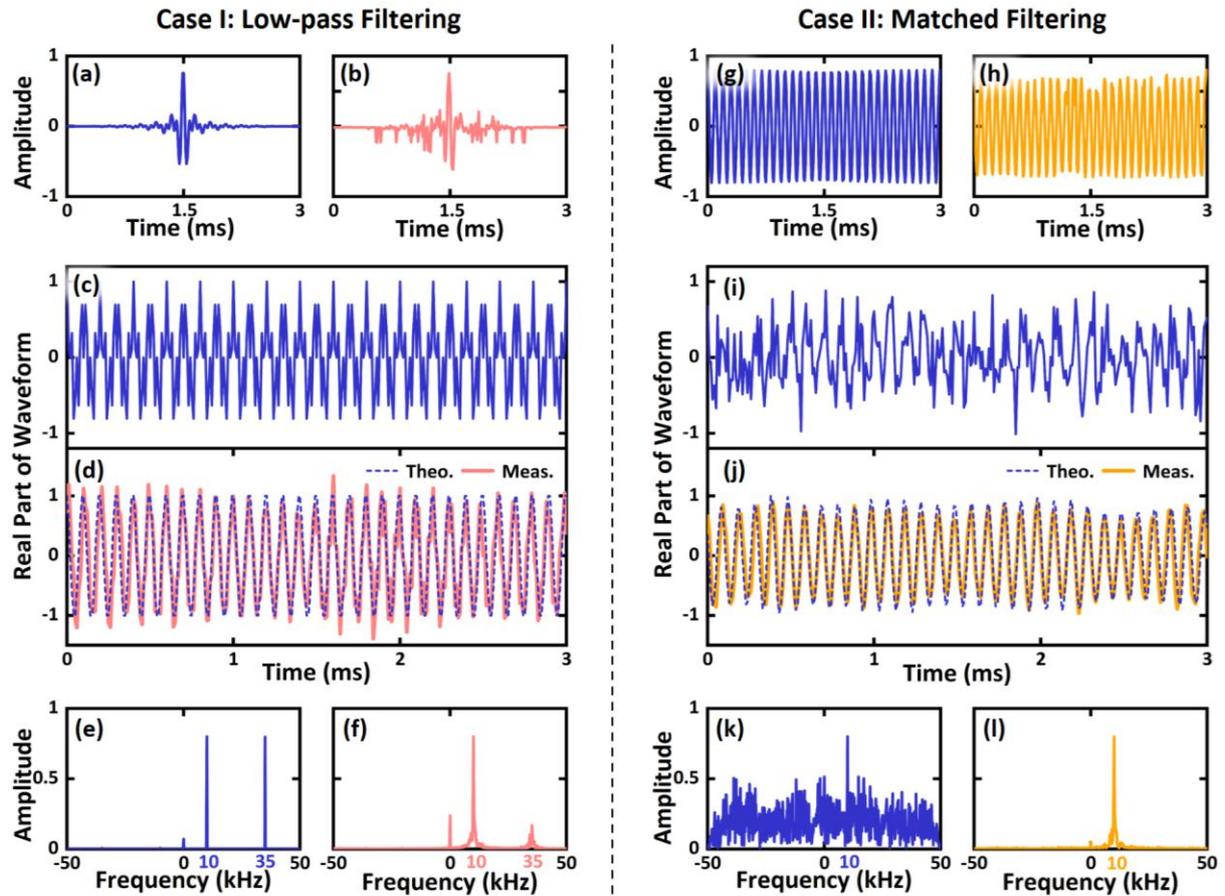

**Fig. 5. Convolution operations performed by the programmable STCM-based RF calculation system, demonstrated for low-pass filtering and matched filtering. (a-f) The low-pass filtering case:** (a, b) Theoretical and measured reflection-coefficient (kernel) waveforms of the STCM. (c) Input signal containing two frequency components. (d) Output waveform compared with theoretical convolution result. (e, f) Amplitude spectra of the input and output signals, showing suppression of the higher-frequency component. **(g-l) The matched filtering case:** (g, h) Theoretical and measured (real part) reflection-coefficient waveforms of the STCM. (i) Input signal: a 10-kHz tone embedded in 0-dB-SNR noise. (j) Output waveform compared with theoretical matched-filter response. (k, l) Spectra of the input and output signals, showing enhancement of the 10-kHz tone and noise suppression.

Table 2. Mathematical models and parameter specifications of two test waveforms for the convolution operation

| Case | | Mathematical model | Parameters |
|---|---|---|---|
| [1] | Input | $x_4[n] = \exp\left\{j \cdot 2\pi f_6 \dfrac{n}{f_s}\right\} + \exp\left\{j \cdot 2\pi f_7 \dfrac{n}{f_s}\right\}$ | $(a,b) = (0.2, 0.1)$ $f_6 = 10$ kHz, $f_7 = 35$ kHz |
| | Kernel | $h_4[n] = \mathrm{Hann}\left(\dfrac{n}{L}\right) \cdot \exp\{af_1 \cdot j\pi n\} \cdot \mathrm{sinc}\left\{b \cdot \left(n - \dfrac{L}{2}\right)\right\}$ | |
| [2] | Input | $x_5[n] = \exp\left\{j \cdot 2\pi f_8 \dfrac{n}{f_s}\right\} + r$ | $f_8 = 10$ kHz |
| | Kernel | $h_5[n] = \exp\left\{-j \cdot 2\pi f_8 \dfrac{L-1-n}{f_s}\right\}$ | |

In both cases, STCM can be switched between the Fourier transform and convolution operators entirely in a software-defined manner, without requiring any modifications to the hardware architecture. Moreover, this real-time programmability is extended to the parameter adjustments within each operation. Together, these capabilities underscore the flexibility and scalability of the RF calculation architecture for various RF signal-processing tasks.

## Applications of STCM-based RF calculations in radar

With the capability of implementing the Fourier transform and convolution in the EM-wave domain, we apply STCM in the radar scenario. In modern coherent radar schemes, velocity measurement relies on Fourier transform-based Doppler processing, while range estimation is based on the convolution operation, as exemplified by pulse compression or matched filter.

For the velocity measurement, the radial velocity of a target relative to a radar system generates a Doppler frequency shift in the echo signals, which can be resolved by the Fourier transform, allowing for the determination of the frequency offset between the transmitted and reflected waves. According to the Doppler-velocity mapping relation $f_d = 2v/\lambda$, where $f_d$ is the Doppler frequency, $v$ is the radial velocity, and $\lambda$ is the wavelength, this frequency offset can be directly converted into target velocity. The detailed theoretical derivation is provided in Supplementary Information 4. In this work, the STCM performs spectrum analysis directly in the EM-domain, enabling the real-time velocity extraction.

In experiments, a 5GHz CW signal was transmitted toward a 200 m/s target synthesized by a radar-echo simulator. The echo signal was incident on the STCM, where the Fourier transform operation was applied directly in the EM-wave domain. As shown in Fig. 6(a), the extracted velocity was 201 m/s, with a 1 m/s deviation from the true value. With 500 sampling points, the theoretical velocity resolution is ±1.5 m/s according to Eq. (S9), indicating that the observed deviation falls within the expected limit. To further improve accuracy, the spectral resolution should be enhanced by increasing the number of sampling points. For example, when the number of samples was increased from 500 to 1000, the theoretical velocity resolution tightened to ±0.75 m/s, and the velocity recalculated by the STCM became 199.5 m/s (shown in Fig. 6(b)), effectively halving the measurement error compared with the previous configuration. Furthermore, to evaluate the robustness of the RF calculation-based velocity measurement system, the targets were configured at 50 m/s intervals in the range from 50 to 300 m/s. Fig. 6(c) demonstrates the measured results and the corresponding error curves, where all velocity errors lie within the resolution limit, validating the effectiveness of the RF calculation approach for radar velocity measurement.

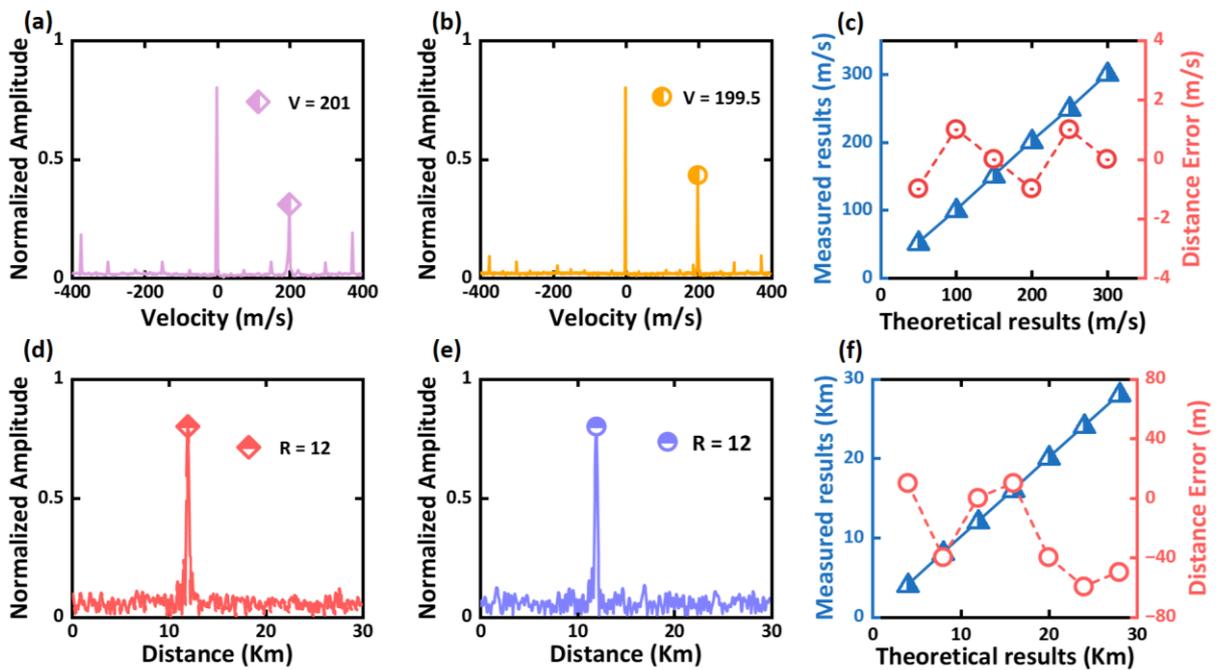

**Fig. 6. Radar detection using the STCM-based programmable RF calculation system.** (a–c) Velocity estimation via RF-domain Fourier transform: (a, b) Measured velocity estimates of 201 m/s and 199.5 m/s obtained with 500 and 1,000 sampling points, respectively (increased points enhance frequency resolution). (c) Measured velocity results and corresponding error curves across 50–300 m/s (in steps of 50 m/s). (d–f) Range estimation via RF-domain Convolution operation: (d, e) Measured range profiles for a target at a true range of 12 km, showing estimated ranges processed using an LFM pulse and a Gold-code sequence waveform, respectively (both with 500 kHz bandwidth, yielding a theoretical range resolution of 300 m). (f) Measured range results and error curves over distances from 4 to 28 km (in steps of 4 km).

For the range measurement, the radar systems commonly employ the pulse compression technique to enhance range resolution. It works by convolving the received echo with a reference waveform that matches the transmitted signal. As a result, the output signal will exhibit peaks whose positions in time correspond directly to the target distances. In our work,

STCM inherently performs this convolution in the EM-wave domain, such that the received signal is processed as it reflects from the metasurface. Consequently, the target ranges can be extracted directly from the reflected waveform without additional electronic processing (Supplementary Information 5). To demonstrate generality, we evaluated two representative radar waveforms: linear frequency-modulated (LFM) pulses and Gold-code sequences, both with a 5 GHz carrier and 500 kHz bandwidth, yielding a 300 m theoretical range resolution. The system sampled each waveform at 5 MHz over a 300 us pulse (1,500 samples). The STCM was reconfigured in a software-defined way to implement the waveform-specific convolution kernel.

In experiments, the echoes from a 12km target were fed into STCM. Accordingly, STCM was configured with kernel-specific coding sequences for the (LFM) pulses and Gold-code sequences, successively. The convolution outputs shown in Fig. 6(d) for the LFM pulses and Fig. 6(e) for the Gold-code sequences exhibit main-lobe peaks at the theoretical target ranges, confirming the accurate programmable pulse compression. Single-target tests from 4 to 28 km in 4 km steps further demonstrate the robustness, with range errors in Fig. 6(f) remaining within the 300 m resolution limit. These results confirm that the STCM-based RF calculation system can handle diverse radar waveforms in a programmable manner, achieving high- precision range extraction without requiring hardware modification.

## Discussion

In this work, we proposed a programmable RF signal processing architecture based on STCM. This architecture can directly perform the Fourier transform and convolution operation of signals in the EM-wave domain, avoiding the limitations of the traditional RF circuits. The programmable STCM-based RF calculation architecture demonstrates three core advantages: programmable capability, exceptional calculation accuracy, and significantly simplified RF hardware through elimination of the cross-domain signal conversion. We firstly implemented two foundational test cases to validate the feasibility of programmable RF calculation principles and calculation accuracy. Building upon this, we successfully implemented real-time echo signal processing using STCM in radar detection scenarios, directly extracting the target range and velocity information to demonstrate the system's application potential. This methodology possesses significant capacity for future applications across radar detection, wireless communications, and intelligent sensing domains.

# Methods

### Details on STCM

The STCM sample is operated at a central frequency of 5 GHz, comprising an 8×12 array of reflective meta-atoms. The sample was simulated using CST Microwave Studio 2020 and fabricated via standard printed circuit board technology. As depicted in Fig. 3(a), the STCM structure features a three-layer configuration for each meta-atom: a top metal patch layer, a middle F4B ($\varepsilon_r = 2.65, \tan\delta = 0.001$) dielectric layer, and a bottom metal ground plane. The reflection coefficient of the meta-atom is adjusted by regulating the on-off states of two integrated PIN diodes on the patch layer. In the simulations, the diode can be modeled as a capacitor-resistor series circuit, as documented in refs[36,50].

### Measurement setups

Fig. 3(d) illustrates the experimental setup used to validate the proposed system's capability for implementing the Fourier transform and convolution operation in the EM-wave domain. A pair of linearly polarized transmitting horn antennas, connected to a vector signal transceiver (NI PXIe-5841), was employed to emit incident signals for the programmable RF calculation system and receive its output information. The state of the STCM sample was controlled by the back-end connected FPGA, I/O interface, and DC power supply modules. Additionally, a radar echo simulator was used to generate radar echoes containing complex target and environmental information, emulating realistic radar measurement environments. The experiment was conducted at 5 GHz, with the spherical coordinate positions (distance, elevation angle, azimuth angle) of the transmitting and receiving antennas set as (1 m, 20°, 0°) and (1 m, −20°, 0°), respectively, with the center of the metasurface array taken as the coordinate origin.

## Author contributions

S.N.C. and Z.Y.C. conceived the project. S.N.C., B.R., and Z.J.Q. performed the experiments. Z.X.W. designed the STCM samples. S.N.C. and J.F.K. wrote the manuscript. Z.Y.C., S.R.W., J.Y.D., Q.C., and T.J.C. revised the manuscript. Q.C. and T.J.C. supervised the project. All authors reviewed and approved the final manuscript.

## Competing interests

The authors declare that they have no competing interests.

## Data availability

The authors declare that all relevant data are available in the paper and its Supplementary Material Files, or from the corresponding author on request.

## Code Availability

The custom computer codes utilized during the current study are available from the corresponding authors on request.

## Acknowledgments

This work is supported by the National Key Research and Development Program of China (2023YFB3811502, 2024YFB2907800), the National Science Foundation (NSFC) for Distinguished Young Scholars of China (62225108), the Fundamental Research Funds for the Central Universities (2242022k60003), the National Natural Science Foundation of China (62288101, 62201139), the Jiangsu Province Frontier Leading Technology Basic Research Project (BK20212002), the Jiangsu Provincial Scientific Research Center of Applied Mathematics (BK20233002), the Jiangsu Science and Technology Research Plan (BK20243028), the Independent Research Fund of the State Key Laboratory of Millimeter Waves (Grant No. Z202503-06), the Fundamental Research Funds for the Central Universities (2242024RCB0005), and the 111 Project (111-2-05).